\documentclass{article}

\PassOptionsToPackage{numbers, compress}{natbib}

\usepackage[final]{tackling_climate_workshop_style}

\usepackage[utf8]{inputenc} 
\usepackage[T1]{fontenc}    
\usepackage{hyperref}       
\hypersetup{hidelinks}
\usepackage{url}            
\usepackage{booktabs}       
\usepackage{amsfonts}       
\usepackage{nicefrac}       
\usepackage{microtype}      

\usepackage{multirow}
\usepackage{enumitem}
\usepackage{amsmath}
\usepackage{graphicx}
\usepackage{svg}

\usepackage{algorithm}
\usepackage{algpseudocode}

\usepackage{subcaption}
\usepackage{booktabs}

\usepackage{pdflscape}

\usepackage{siunitx} 
\newcommand{\aifstc}{AIFS--TC}
\newcommand{\aifstcsat}{AIFS--TCSAT}

\title{AIFS--TCSAT: Causal Forecast Corrections from Intermittent Satellite Observations}
\author{Pritthijit Nath\\
Department of Applied Mathematics and Theoretical Physics, University of Cambridge\\
\texttt{pn341@cam.ac.uk}}

\begin{document}
\setcitestyle{square}
\raggedbottom
\maketitle

\begin{abstract}
Satellite products provide valuable information on tropical-cyclone (TC) inner-core structure, but their arrival times vary across observing systems. Retrospectively incorporating these observations can improve forecasts while inadvertently introducing information unavailable at the original issue time. To preserve this causal availability constraint, we introduce \aifstcsat{}, an event-driven extension of the new AIFS-TC intensity correction that issues an initial forecast from the information available at the time and then produces immutable revisions as GEO, PMW, SAR, and operational-fix observations arrive. In a storm-disjoint retrospective evaluation of 30 North Atlantic and eastern North Pacific cyclones from 2023, the full multimodal correction reduces storm-balanced wind MAE by 6.3\% relative to the AIFS-TC baseline and 5.2\% relative to fix-only correction. These results show promise for correction mechanisms that incorporate newly available satellite observations to refine TC intensity forecasts in future forecast settings.
\end{abstract}

\section{Introduction}

Since 2023, global artificial intelligence (AI) weather models~\cite{lam_learning_2023,bi_accurate_2023,price_probabilistic_2025} have enabled cheap and skilful large-scale forecasts, yet their coarse grids and objectives can smooth the compact inner-core processes governing tropical cyclone (TC) intensity. Here, we focus on ECMWF's operational deterministic model AIFS-Single~\cite{lang_aifs_2024}, trained on ERA5 reanalysis~\cite{hersbach_era5_2020} to produce 15-day global forecasts. For correcting TC intensity biases, \aifstc{}~\cite{allen_aifstc_2026} combines gradient-boosted adjustments from scalar predictors with convolutional corrections from storm-centred AIFS fields. However, these predictors remain derived primarily from the forecast model and may not fully capture observed inner-core structure. Satellite observations provide a complementary source of information on these features, making them particularly relevant for rapid intensification (RI), defined as an increase in maximum sustained 10-m wind of at least 30~kt within 24~h~\cite{kaplan_largescale_2003}. This challenge is likely to become increasingly important under anthropogenic warming, as the probability of RI is expected to increase with more favourable oceanic and atmospheric conditions for intensification~\cite{bhatia_potential_2022}.

Event-driven extensions to such ML systems are not equivalent to simple additions of synchronous input channels. Satellite modalities including geostationary (GEO) and passive-microwave (PMW) imagery, synthetic-aperture-radar (SAR) surface winds, and operational fixes arrive at different cadences. Retrospectively selecting the ``latest'' observation can therefore introduce information unavailable at forecast issue time, while a missing modality may reflect observing-system sampling rather than a zero-valued field. In this study, we therefore investigate whether asynchronous events can causally revise an initial forecast and whether satellite inputs improve matched revisions relative to operational fixes alone. 

\clearpage

The key contributions of this work are:
\begin{enumerate}[leftmargin=*,nosep]
    \item \textbf{Causal intermittent (CI) updates}: We introduce event-driven updates that admit only information available by reissue time while preserving each revision as an immutable and auditable forecast.
    \item \textbf{Matched evaluation}: We compare satellite-plus-fix and fix-only corrections under identical storms, forecasts, targets, baselines, and optimisation limits.
    \item \textbf{Improved TC forecasts}: The full extension reduces storm-balanced wind and RI-positive errors by 6.3\% and 6.2\% relative to \aifstc{}, while pressure and RI-probability results remain mixed.
\end{enumerate}

\section{Methodology}

\subsection{Model hierarchy}
The model hierarchy comprises raw AIFS-Single, a frozen \aifstc{} baseline, a fix-only correction using later operational estimates, and the full correction adding GEO, PMW, and SAR events. Following Allen et al.~\cite{allen_aifstc_2026}, our clean-room reconstruction blends gradient-boosted scalar and convolutional storm-centred AIFS corrections with the raw intensity forecast. Same-cycle CARQ wind and pressure estimates are included only in the frozen \aifstc{} baseline, while the updater receives later CARQ estimates. The fix-only and full corrections therefore share the same baseline and operational-fix pathway, isolating the contribution of GEO, PMW, and SAR observations.

\subsection{Data processing}
\label{sec:methods-data-processing}
The preliminary experiment includes 30 complete 2023 storms from the North Atlantic and eastern North Pacific, covering waters off the US East and West coasts. AIFS-Single hindcasts are initialised from ERA5~\cite{hersbach_era5_2020}, while storm identity and verification intensity come from IBTrACS v4r01~\cite{gahtan_ibtracs_2024}. Best-track data are used only for targets and evaluation and excluded from inference, centring, normalisation, and calibration. CARQ operational fixes reconstructed from ATCF A-decks~\cite{nrl_atcf_2026} form a separate timestamped stream. Satellite events comprise GOES R-series ABI imagery~\cite{noaa_goescmi_2026}, TC PRIMED microwave brightness temperatures~\cite{razin_tcprimed_2023}, and CyclObs SAR surface winds~\cite{mouche_fostering_2026}. Further dataset details are provided in \ref{sec:appendix-addln-dataset-details}.

Five outer folds partition storms before crops, forecast windows, or preprocessing statistics are constructed, preventing any storm from contributing to both fitting and held-out evaluation. This yields 236 AIFS initialisation-storm forecast tasks (inventory creation details in \ref{sec:appendix-forecast-inventory-creation}) and 13,125 storm-centred satellite crops. Replaying each task as admissible events arrive produces 14,511 immutable forecast snapshots and 30,909 target-resolved matched forecasts, since each snapshot can be evaluated at multiple valid times.

\subsection{Causal intermittent (CI) updates}

Causal intermittent (CI) updates refer to event-driven forecast revisions that admit only observations available by the reissue time. Given a frozen \aifstc{} baseline $b_0$ issued at initial time $t_0$, each event $e_j$ has a recorded availability time $\tau_j^{\mathrm{avail}}$, and the admissible event set at reissue time $t$ is $\mathcal{E}(t)=\{e_j:\tau_j^{\mathrm{avail}}\leq t\}$. For a forecast lead $\ell$-timesteps measured from the reissue time,
\begin{equation}
 \widehat{\mathbf y}(t,\ell)=\widehat{\mathbf y}_0(b_0,\ell)
 +g_\theta\!\left(b_0,\ell,\mathcal{E}(t)\right),
 \label{eq:ci-update}
\end{equation}
where $\widehat{\mathbf y}(t,\ell)$ is the revised maximum-wind and minimum-pressure forecast, $\widehat{\mathbf y}_0(b_0,\ell)$ is the corresponding frozen baseline forecast, and $g_\theta$ is the learned correction with parameters $\theta$. The correction receives baseline features $b_0$, lead $\ell$, and the admissible event set $\mathcal{E}(t)$, and returns bounded wind and pressure residuals together with an RI logit. The initial forecast uses only events available at $t_0$. Each future arrival at lead $\ell$ creates a new immutable snapshot from $b_0$ and $\mathcal{E}(t)$ using Equation~\ref{eq:ci-update}.

\begin{figure}[!h]
 \centering
 \vspace{-1.25cm}
 \makebox[\textwidth][c]{\includegraphics[width=1.4\linewidth]{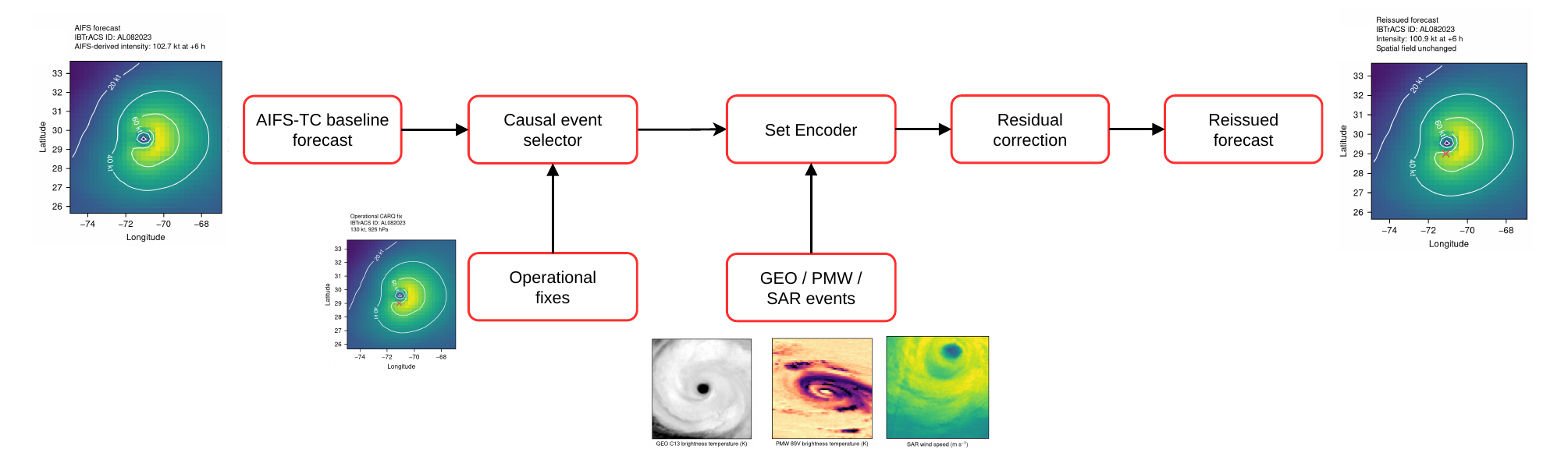}}
 \caption{High-level causal forecast--revision cycle. Only events available by issue time $t$ are admitted to the set encoder. The correction revises TC intensity estimates rather than the atmospheric state and therefore does not constitute dynamical data assimilation.}
 \label{fig:architecture}
 \vspace{-2mm}
\end{figure}

Figure~\ref{fig:architecture} summarises the update architecture. Satellite images and operational fixes are encoded into modality-specific event representations containing timing, quality, coverage, identity, and presence information. A permutation-invariant set~\cite{zaheer_deepsets_2017} operation pools any number of events into a single update representation, with duplicate events idempotent and empty sets returning the baseline exactly. Recorded availability timestamps are used to enforce forecast-time causality, while acquisition-time replay is explicitly labelled retrospective whenever true historical dissemination times are unavailable.

\clearpage

\subsection{Matched evaluation}

\enlargethispage{2\baselineskip}

The full and fix-only systems use identical storms, folds, episodes, issue and valid times, targets, baselines, and optimisation limits, with one model trained per outer fold for each system (as described in Section~\ref{sec:methods-data-processing}). As measures of TC intensity, maximum sustained wind and minimum central pressure are evaluated, although no physical relationship between them is enforced. For continuous intensity metrics and RI probability, we use storm-balanced MAE and Brier score respectively, averaging first within each storm  before assigning equal weight across the $S$ storms:

\begin{figure}[H]
\centering
\begin{minipage}{0.48\textwidth}
\vspace{-5mm}
\begin{equation}
\mathrm{MAE}_{m}
=
\frac{1}{S}\sum_{s=1}^{S}
\frac{1}{n_s}\sum_{i=1}^{n_s}
\left|\widehat{y}_{s,i,m}-y_{s,i,m}\right|
\label{eq:storm-balanced-mae}
\end{equation}
\end{minipage}
\hfill
\begin{minipage}{0.48\textwidth}
\begin{equation}
\mathrm{BS}
=
\frac{1}{S}\sum_{s=1}^{S}
\frac{1}{n_s}\sum_{i=1}^{n_s}
(p_{s,i}-o_{s,i})^2
\label{eq:brier-score}
\end{equation}
\end{minipage}
\vspace{-3mm}
\end{figure}

Here, $n_s$ is the number of eligible forecasts for storm $s$, while $p_{s,i}$ is the predicted RI probability and $o_{s,i}\in\{0,1\}$ indicates whether RI occurred. RI-positive wind MAE applies Equation~\ref{eq:storm-balanced-mae} only to forecasts whose verifying 24-h wind change meets the 30-kt RI threshold and then averages equally across storms containing eligible forecasts. Uncertainty is estimated from 10,000 paired bootstrap replicates resampling complete storms within outer-fold and basin strata. Pairing preserves matched forecasts, while whole-storm resampling retains within-cyclone dependence. Reported 95\% intervals reflect storm-sampling uncertainty conditional on the fitted model with negative full-minus-fix-only differences favouring the full model.

\section{Results}
On the 1,154-prediction baseline-construction population, \aifstc{} reduces wind MAE from 13.963 to 10.232~kt relative to raw AIFS, a 26.7\% reduction (not shown). The underlying AIFS forecasts are retrospective AIFS-Single hindcasts initialised from ERA5. On 30,909 matched held-out forecasts derived from 14,511 immutable reissue snapshots, the full model (shown in Table~\ref{tbl:results-model-scores}) reduces wind MAE from 9.927 to 9.410~kt relative to fix-only, a 5.2\% reduction. The paired difference is $-0.516$~kt (95\% interval $[-0.867,-0.175]$~kt), resolving a bounded improvement under whole-storm resampling.

\begin{table}[!h]
\centering
\caption{Model scores for the 30 complete storms in 2023. Comparisons between the frozen \aifstc{} baseline and both corrections on the same 30,909 matched forecasts. Brackets show 95\% whole-storm bootstrap intervals.} \vspace{2mm}
\label{tbl:results-model-scores}
\small
\setlength{\tabcolsep}{3pt}
\makebox[\linewidth][c]{%
\begin{tabular}{llcccc}
\toprule
Model & Wind MAE (kt) $\downarrow$ & Pressure MAE (hPa) $\downarrow$ & RI Brier score $\downarrow$ & RI-positive wind MAE (kt) $\downarrow$ \\
\midrule
\aifstc{} baseline & \shortstack{10.044\\{[7.274, 13.229]}} & \shortstack{\textbf{6.478}\\{[4.384, 8.877]}} & \shortstack{0.0767\\{[0.0209, 0.1449]}} & \shortstack{14.485\\{[7.651, 21.524]}} \\
+ Fix-only correction & \shortstack{9.927\\{[7.185, 13.066]}} & \shortstack{6.481\\{[4.386, 8.879]}} & \shortstack{\textbf{0.0746}\\{[0.0296, 0.1258]}} & \shortstack{14.094\\{[7.257, 21.149]}} \\
+ Full correction & \shortstack{\textbf{9.410}\\{[6.826, 12.388]}} & \shortstack{6.629\\{[4.579, 8.964]}} & \shortstack{0.0837\\{[0.0279, 0.1517]}} & \shortstack{\textbf{13.590}\\{[6.758, 20.490]}} \\
\bottomrule
\end{tabular}
}
\end{table}

\clearpage

\begin{figure}[t]
 \centering
 \includegraphics[width=0.75\linewidth]{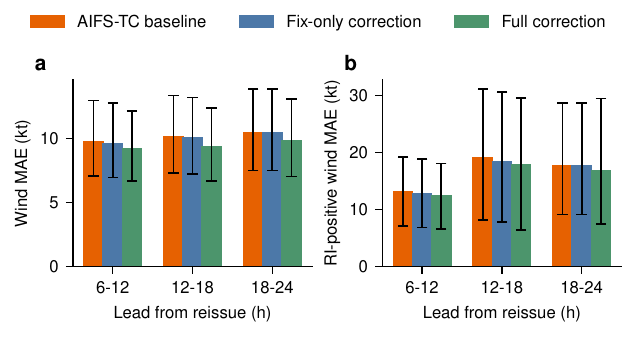}
 \caption{(a) Wind MAE by lead from reissue for all held-out forecasts  and (b) RI-positive queries. Orange, blue, and green bars denote the frozen \aifstc{} baseline, fix-only, and full corrections, respectively. Whiskers show 95\% whole-storm bootstrap intervals stratified by outer fold and basin. The bins contain 12,897, 10,360, and 7,652 matched forecasts. AIFS forecasts follow 6-hr intervals.}
 \label{fig:wind-mae-bar-whiskers}
 \vspace{-4mm}
\end{figure}

Figure~\ref{fig:wind-mae-bar-whiskers}a disaggregates wind accuracy by lead from reissue. Full-correction wind MAE is 9.203, 9.341, and 9.837~kt at 6--12, 12--18, and 18--24~h, compared with 9.609, 10.026, and 10.446~kt for fix-only, corresponding to reductions of 4.2\%, 6.8\%, and 5.8\%, respectively. The binwise summaries are descriptive rather than separate hypothesis tests, but their consistent direction shows that the aggregate wind result is not confined to one reported lead bin.

The RI diagnostics in Figure~\ref{fig:wind-mae-bar-whiskers}b are little uncertain. On RI-positive queries, full-correction wind MAE is 12.366, 17.954, and 16.768~kt across the three lead bins, compared with 12.808, 18.437, and 17.682~kt for fix-only. Across all RI-positive queries, the full correction reduces wind MAE by 3.6\%, from 14.094 to 13.590~kt, but its paired 95\% interval includes zero. Only seven storms contribute RI-positive queries, which limits the precision of this conditional comparison.

The remaining mean changes in Table~\ref{tbl:results-model-scores} do not support improvements under whole-storm resampling. Full-correction pressure MAE is 2.3\% higher than fix-only, at 6.629 versus 6.481~hPa, while its 95\% paired interval includes zero. The RI Brier score is 12.2\% higher, at 0.0837 versus 0.0746, and its interval also includes zero. These outcomes support improved maximum-wind and RI-positive wind accuracy across all matched forecasts, but not improved central pressure or RI probability.

\section{Discussion and conclusion}

\aifstcsat{} demonstrates that irregular observations can revise an AIFS-derived TC intensity forecast while preserving a reproducible forecast-time boundary. In the baseline construction, the frozen \aifstc{} reconstruction reduces raw-AIFS wind MAE by 26.7\%. In the separate matched-correction population (shown in Table~\ref{tbl:results-model-scores}), the full correction reduces wind MAE by 6.3\% relative to the frozen \aifstc{} baseline and by 5.2\% relative to the fix-only correction.

Because these populations differ, the full correction cannot be compared directly with raw-AIFS scores, and results remain conditional on retrospective AIFS hindcasts initialised from ERA5. The joint satellite pathway adds predictive information beyond operational fixes for this fitted 2023 model, but does not isolate individual sensor contributions, enforce a physically coherent wind--pressure state, or demonstrate operational skill. The single-season, single-fit design also leaves temporal and optimisation uncertainty unmeasured. Other identified limitations are mentioned in \ref{sec:appendix-limitations}.

Future experiments underway will use multi-year temporal training and holdout with measured product receipt times. They will prespecify RI analyses, perform modality-removal retraining, quantify optimisation sensitivity, and add RI calibration diagnostics. These tests will assess whether the wind gains persist across different hold years and which observation pathways drive them. The present study provides a causal and auditable framework for testing satellite-based forecast corrections and lays the foundation for their evaluation in future real-time operational settings.

\clearpage

\begin{ack}
P. Nath was supported by the UKRI Centre for Doctoral Training in Application of Artificial Intelligence to the study of Environmental Risks (EP/S022961/1). The author acknowledges the use of resources provided by the Isambard-AI National AI Research Resource (AIRR). Isambard-AI is operated by the University of Bristol and is funded by DSIT via UKRI and the STFC [ST/AIRR/I-A-I/1023]. The author gratefully acknowledges Dr Anna Jungbluth (Imperial College London) and the FDL ESL 2026 Tropical Cyclone Dynamics team for allowing the reuse of the dataset curated for the challenge.
\end{ack}

\section*{LLM usage disclosure}
The author acknowledge the use of AI language models, specifically OpenAI Codex and ChatGPT~(GPT-5.6 Sol and GPT-6 Astra), during the preparation of this work. These tools were used to assist in software development, polish language usage and improve the overall clarity of the manuscript. All AI-generated content was reviewed, verified, and edited by the author to ensure accuracy and appropriateness.

\small
\bibliographystyle{vancouver}
\bibliography{bibliography}

\normalsize

\clearpage
\appendix
\renewcommand\thesection{Appendix \Alph{section}}
\renewcommand\thesubsection{\Alph{section}.\arabic{subsection}}
\renewcommand\thetable{\Alph{section}.\arabic{table}}
\renewcommand\thefigure{\Alph{section}.\arabic{figure}}
\setcounter{table}{0}
\setcounter{figure}{0}

\clearpage
\begin{landscape}
\section{Additional dataset details}
\label{sec:appendix-addln-dataset-details}

\begin{table}[!ht]
\centering
\caption{Dataset variables, use, and access information for the preliminary 2023 experiment.} \vspace{2mm}
\label{tab:appendix-datasets}
\footnotesize
\begin{tabular}{r >{\raggedright\arraybackslash}p{0.178\linewidth}>{\raggedright\arraybackslash}p{0.105\linewidth}>{\raggedright\arraybackslash}p{0.145\linewidth}>{\raggedright\arraybackslash}p{0.205\linewidth}>{\raggedright\arraybackslash}p{0.237\linewidth}}
\toprule
& Dataset & Product type & Variables & Use & Source \\
\midrule
1 & European Centre for Medium-Range Weather Forecasts (ECMWF) Reanalysis v5 (ERA5)~\cite{hersbach_era5_2020} & Global atmospheric reanalysis & Atmospheric fields used by the forecast model & Initialises the Artificial Intelligence Forecasting System (AIFS)-Single hindcasts from which the frozen forecast baseline is constructed & \url{https://doi.org/10.1002/qj.3803} \\ \midrule
2 & International Best Track Archive for Climate Stewardship (IBTrACS), version 4r01~\cite{gahtan_ibtracs_2024} & Best-track archive & Storm identity, maximum sustained wind, and minimum central pressure & Provides targets and evaluation data; excluded from inference, crop centring, normalisation, and calibration & \url{https://doi.org/10.25921/82ty-9e16} \\ \midrule
3 & Combined Automated Request Query (CARQ) records from Automated Tropical Cyclone Forecasting System (ATCF) A-decks~\cite{nrl_atcf_2026} & Operational-fix archive & Storm position and intensity estimates with record times & Supplies correction inputs and causal satellite-crop centres; only records available no later than the event time are admitted & \url{https://science.nrlmry.navy.mil/atcf/docs/html/ATCF_SAG_Sec2.html} \\ \midrule
4 & Geostationary Operational Environmental Satellite R-series Advanced Baseline Imager (GOES-R ABI)~\cite{noaa_goescmi_2026} & Level-2 geostationary imagery & Multiband cloud and moisture imagery & Constructs storm-centred satellite events for the full correction & \url{https://www.ncei.noaa.gov/products/goes-terrestrial-weather-abi-glm} \\ \midrule
5 & Tropical Cyclone Precipitation, Infrared, Microwave, and Environmental Dataset (TC PRIMED)~\cite{razin_tcprimed_2023} & Passive-microwave archive & Microwave brightness temperatures & Constructs storm-centred satellite events for the full correction & \url{https://doi.org/10.1175/BAMS-D-21-0052.1} \\ \midrule
6 & CyclObs synthetic-aperture-radar (SAR) surface winds~\cite{mouche_fostering_2026} & Processed radar observations & Surface-wind fields & Constructs storm-centred satellite events for the full correction & \url{https://doi.org/10.1016/j.rse.2025.115139} \\
\bottomrule
\end{tabular}
\end{table}
\end{landscape}
\clearpage

\section{Forecast inventory creation}
\label{sec:appendix-forecast-inventory-creation}

\begin{enumerate}[leftmargin=*,label=\arabic*.,itemsep=0.7em]
\item \textbf{Construct the forecast-task cohort:} The forecast-task inventory is constructed from AIFS-Single hindcasts without using IBTrACS best-track data. TC candidates are detected using TempestExtremes~\cite{ullrich_tempestextremes_2017} v2.4.2, which identifies minima in mean sea-level pressure, requires a nearby closed pressure contour and warm-core geopotential-thickness signature, and links candidates through time subject to spatial, duration, wind, and latitude criteria. Each candidate track is associated with a storm only through its position at the AIFS issue time and the corresponding CARQ position, with ambiguous matches rejected. This procedure along with a frozen recovery configuration identifies 236 forecast tasks.

\item \textbf{Build causal crops and data splits:} Every satellite crop is centred on the latest same-storm CARQ position available no later than the event time rather than on a retrospective best-track position. Five complete-storm outer folds are assigned before crop construction and preprocessing, and fold-specific statistics are fitted without held-out storms. Automated checks reject future observations, best-track inference inputs, cross-fold storms, mismatched comparator identities, non-causal centres, and conflicting duplicate artefacts.

\item \textbf{Record and audit each revision:} Each forecast sidecar records the issue and valid times, selected event list, outer fold, seed, model and normalisation hashes, configuration, training and replay episodes, and source manifest. The event encoder is permutation invariant, and duplicate identifiers are removed deterministically, ensuring that reordering or duplicating an event cannot change a prediction. Together, these records make every immutable revision reconstructable and auditable against the availability rule.
\end{enumerate}

\section{Current identified limitations and reproducibility}
\label{sec:appendix-limitations}

\begin{table}[h]
\centering
\caption{Current identified limitations of the preliminary experiment and their implications.} \vspace{2mm}
\label{tab:appendix-limitations}
\small
\begin{tabular}{r p{0.27\linewidth}p{0.59\linewidth}}
\toprule
& Limitation & Implication(s) \\
\midrule
1 & Single 2023 season & Storm-disjoint folds prevent cyclone overlap, but shared observing systems, climate state, processing software, and development decisions preclude claims of temporal generalisation. \\
2 & One prespecified fit & Whole-storm intervals quantify storm-sampling uncertainty conditional on seed 42, but not optimisation uncertainty. \\
3 & Reconstructed availability & Missing historical receipt times prevent measured operational-latency or readiness claims. \\
4 & Non-random sensor coverage & Joint training and uneven sampling prevent causal attribution of the wind result to an individual satellite modality. \\
5 & \aifstc{} baseline & The clean-room reconstruction cannot be compared directly with the independent 2025 evaluation of Allen et al.~\cite{allen_aifstc_2026} because the source code is unavailable (as of 29 August 2026) and the forecast population and evaluation period differ. \\
6 & Intensity-only correction & The correction does not update the spatial atmospheric state or enforce dynamical balance or a wind--pressure relationship. \\
7 & Seven RI-positive storms & RI-positive wind error and RI-probability conclusions remain imprecise and require a larger, prespecified temporal sample. \\
\bottomrule
\end{tabular}
\end{table}

The source code (to be released upon acceptance) includes configurations, tests, a paper-local audited result snapshot, and scripts for regenerating every figure. Exact model architecture will be released with source code upon acceptance. Raw observations remain subject to their source licences.

\end{document}